\documentclass{article}
\usepackage{spconf,amsmath,graphicx}
\PassOptionsToPackage{hyphens}{url}\usepackage{hyperref}

\usepackage{xcolor}
\usepackage{enumitem, paralist}
\usepackage[caption=false]{subfig}
\usepackage{float}

\usepackage{multirow}
\usepackage{makecell}
\usepackage{multicol}
\usepackage{tabularx}
\usepackage{color}
\usepackage{booktabs}
\usepackage[dvips]{epsfig}

\usepackage{appendix}

\usepackage{setspace}
\usepackage[utf8]{inputenc}

\usepackage[
backend=biber,
style=ieee,
maxbibnames=3,
maxcitenames=3,
doi=false,isbn=false,url=false,eprint=false
]{biblatex}
\defbibheading{bibliography}[\refname]{}

\newcommand{\redit}[1]{\textcolor{black}{#1}}

\newcommand{\rredit}[1]{\textcolor{black}{#1}}
\newcommand{\rrredit}[1]{\textcolor{black}{#1}}

\newcommand{\rpostedit}[1]{\textcolor{black}{#1}}

\newcommand\figref[1]{Fig.~\ref{#1}}
\newcommand\tabref[1]{Table~\ref{#1}}
\newcommand\secref[1]{Section~\ref{#1}}

\title{NNSVS: A Neural Network-Based Singing Voice Synthesis Toolkit}
\name{Ryuichi Yamamoto$^{1,2}$, Reo Yoneyama$^2$, and Tomoki Toda$^2$}
\address{$^{1}$LINE Corp., Japan, $^{2}$Nagoya University, Japan.}
\begin{document}
\fontsize{9.2}{11.1}\selectfont
\maketitle
\begin{abstract}
This paper describes the design of NNSVS, an open-source software for neural network-based singing voice synthesis research. NNSVS is inspired by Sinsy, an open-source pioneer in singing voice synthesis research, and provides many additional features such as multi-stream models, autoregressive fundamental frequency models, and neural vocoders. Furthermore, NNSVS provides extensive documentation and numerous scripts to build complete singing voice synthesis systems. Experimental results demonstrate that our best system significantly outperforms our reproduction of Sinsy and other baseline systems. The toolkit is available at \url{https://github.com/nnsvs/nnsvs}.
\end{abstract}
\section{Introduction}
\label{sec:intro}


Open-source software has played a crucial role in advancing research; 
for example, PyTorch~\cite{paszke2019pytorch} and TensorFlow~\cite{tensorflow2015-whitepaper} for deep learning, HTK~\cite{young2002htk} and Kaldi~\cite{povey2011kaldi} for speech recognition, and HTS~\cite{zen2007hmm} and Merlin~\cite{wu2016merlin} for speech synthesis, have been extensively used by the research and industry communities.

Sinsy is an open-source pioneer in singing voice synthesis (SVS)~\cite{oura2010recent,hono2018recent, hono2021sinsy}.
Sinsy has had more than 10 years of development history since its first public release. 
\rpostedit{Sinsy} adopted statistical parametric SVS based on hidden Markov models (HMMs) in the first version and switched to deep neural networks (DNNs) to improve SVS quality.
Although the SVS community greatly benefited from their efforts, the functionality of its open-source version is limited to traditional HMM-based SVS and DNN-based SVS is not publicly available. 


Most recently, a new open-source toolkit for end-to-end SVS, Muskits has been proposed~\cite{shi22d_interspeech}. 
Although Muskits provides several DNN-based SVS models~\cite{lu2020xiaoicesing,blaauw2020sequence,shi2021sequence} \rredit{with a number of reproducible recipes~\cite{povey2011kaldi}}, it does not support \rredit{well-designed} parametric approaches that can achieve both good quality and \rredit{pitch robustness}, as in the latest Sinsy~\cite{hono2021sinsy}.


In this paper, we propose NNSVS, a new open-source toolkit for singing voice synthesis (SVS) written in Python and PyTorch~\cite{paszke2019pytorch}. 
In contrast to Muskits, 
NNSVS does not specialize in end-to-end SVS. Instead, we aim to provide a modular and extensible codebase that is easily applied to various SVS architectures including parametric SVS~\cite{blaauw2017neural,hono2018recent,lu2020xiaoicesing}, modern SVS using neural vocoders~\cite{gu2021bytesing,liu2021diffsinger}, and their hybrid methods~\cite{yi2019singing,hono2021sinsy}. 
The important features are summarized as follows:

\textit{Modular design}: Following Sinsy's structure~\cite{hono2021sinsy}, NNSVS decomposes an SVS system into four core modules: the time-lag model, duration model, acoustic model, and vocoder.

\textit{Extensible design}: Together with the modular design, every module can be flexibly \rredit{customized}.
\rredit{
For example, users can add new acoustic model architectures without modifying the remaining modules.
Furthermore, our generic multi-stream implementation of acoustic models gives users fine-grained control over the model architecture for each feature stream separately.
}

\textit{Language-independent design}: The \rredit{above-mentioned four} core modules are language-independent by design. Therefore, users can create SVS systems for custom languages by implementing a language-dependent pre-processing (e.g., extracting phonetic contexts from musical scores). 

\textit{Everything is open-source}: In contrast to the open-source version of Sinsy, our code is fully open-sourced. We also provide an implementation that resembles Sinsy as a baseline system \rredit{to further encourage reproducible research}.

\textit{Complete recipes}: Following the success of Kaldi~\cite{povey2011kaldi}, ESPnet~\cite{watanabe2018espnet}, and Muskits~\cite{shi22d_interspeech}, we provide complete setups for building SVS systems.

\textit{Documentation}: NNSVS is extensively documented. Documentation is available online\footnote{\url{https://nnsvs.github.io/}}.

Even though the modular and extensible design allows users to implement their custom models and recipes, our toolkit provides several baseline implementations.
In particular, we provide varieties of acoustic models such as those of Sinsy~\cite{hono2018recent,hono2021sinsy}, multi-stream models~\cite{blaauw2017neural,kim2018korean}, and  autoregressive fundamental frequency ($F_0$) models~\cite{yi2019singing}.
Furthermore, to obtain the best performance possible of traditional parametric approaches and \rrredit{recent} neural vocoders, we incorporate the unified \rredit{source-filter} generative adversarial networks (uSFGAN)~\cite{yoneyama22_interspeech} to achieve high-quality and \rredit{pitch-robust} SVS systems.

To evaluate the quality of the SVS systems, we compare NNSVS with \rredit{some baseline systems including} Muskits~\cite{shi22d_interspeech}, Sinsy~\cite{hono2021sinsy}, and a recently proposed modern SVS system called DiffSinger~\cite{liu2021diffsinger}.
Experimental results demonstrate that our best system archives a mean opinion score (MOS) of 3.86, significantly outperforming \rredit{the} baseline systems.


\section{Data representation}
\label{sec:data}

To build SVS systems with NNSVS, the following data are required: 1) the waveform, 2) musical score, and 3) phone segmentation. The latter two can be represented as an HTS label file~\cite{zen2007hmm}, which includes timings (i.e., the start and end time of each phone) and phonetic/musical contexts.

\subsection{Musical score features}
Inspired by Sinsy~\cite{hono2021sinsy}, our toolkit uses HTS full-context labels as the primary musical score representation.
We provide tools to obtain HTS labels from MusicXML~\cite{good2001musicxml} and UST\footnote{
Files created by vocal synthesizer UTAU~\cite{utau} and its open-source successor OpenUTAU~\cite{openutau}.
}~\cite{shen2022linguistic} files.
We also provide a set of context definitions \rredit{(i.e., phonetic and musical contexts to be extracted from HTS labels)} designed for Japanese SVS.  
\rredit{Note that} they can be extended to \rredit{other} languages (e.g., UK, US, or Australian English~\cite{nnsvs_english}).
Furthermore, \rredit{to control the characteristics of the synthetic voice}, users can add custom contexts such as falsetto flags, emotion flags, and strength of voice.

Given the context definitions, our toolkit converts HTS labels to \rredit{note-level and phone-level} musical score features that consist of categorical (e.g., phone identity) and numeric features (e.g., note pitch and duration).
Then, those extracted features are used as inputs for the neural networks.

\subsection{Acoustic features}

We use WORLD~\cite{Morise2016WORLDAV} as the main \rredit{acoustic} feature extraction tool. 
WORLD decomposes audio signals into $F_0$, the spectral envelope, and band-aperiodicity (BAP). 
The spectral \rredit{envelope is} converted to \rredit{mel-generalized} cepstral coefficients (MGCs) to reduce dimensionality without quality deterioration~\cite{morise17_interspeech}.
Furthermore, $F_0$ is converted into a continuous log-scale $F_0$ (log-$F_0$)~\cite{yu2010continuous} \rredit{and voiced/unvoiced flags (VUVs)} for the ease of modeling.

The extracted acoustic features are used as the target features of the acoustic models and conditioning features for the neural vocoders.
We also integrate mel-spectrogram feature extraction to support recent SVS model architectures that predict mel-spectrograms from the musical score.
To support explicit vibrato modeling proposed in Sinsy~\cite{hono2021sinsy}, we provide optional vibrato parameter extraction~\cite{nakano2006automatic}.


\section{Core modules}
\label{sec:core}

\begin{figure}[!t]
\begin{minipage}[t]{.5\linewidth}
\hspace*{3mm}
\centerline{\epsfig{figure=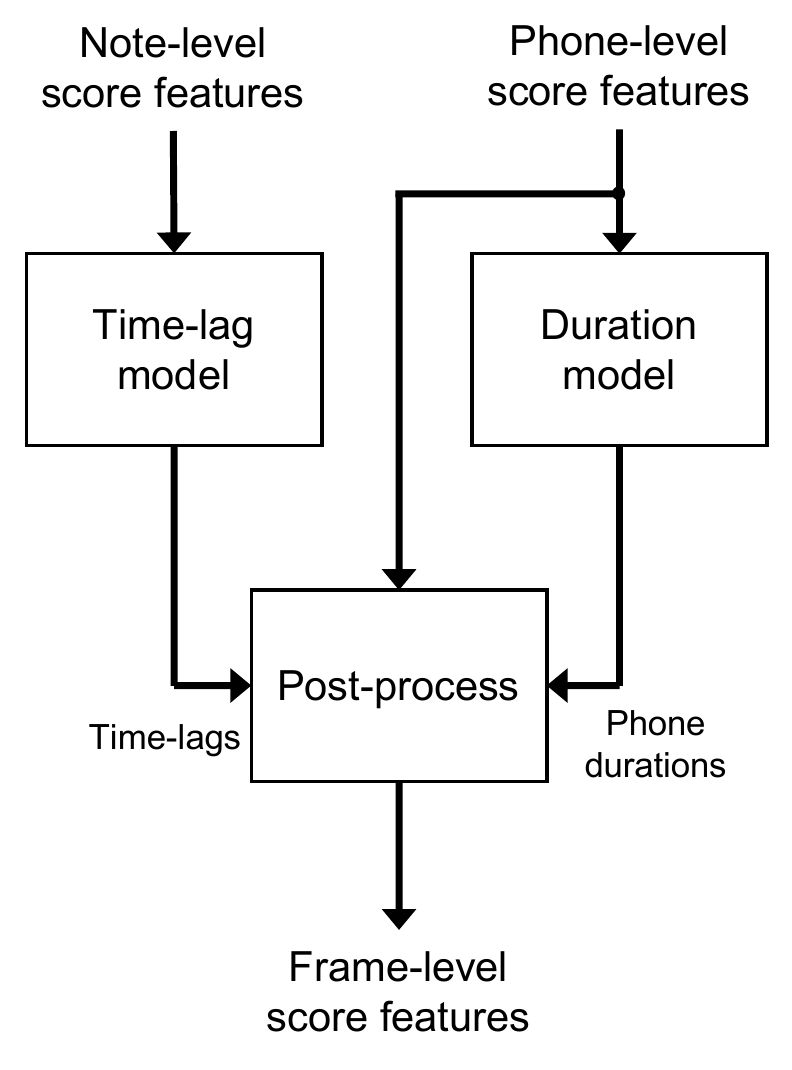,height=55mm}}
\hspace*{3mm}
\centerline{(a)}  \medskip
\end{minipage}
\begin{minipage}[t]{.5\linewidth}
\hspace*{1mm}
\centerline{\epsfig{figure=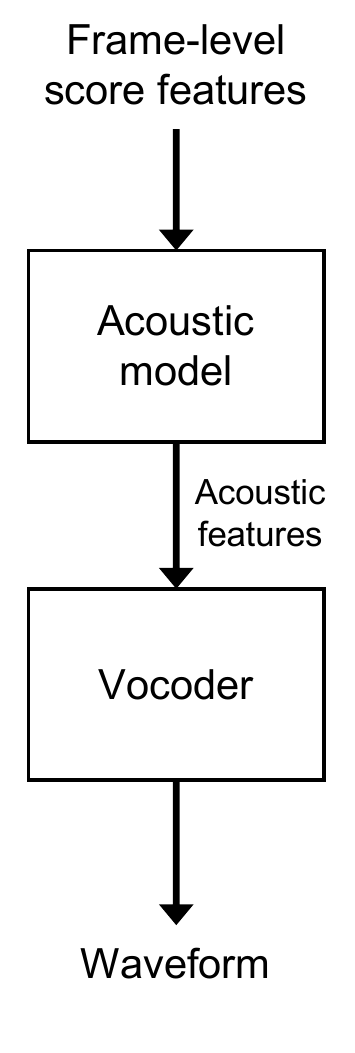,height=55mm}}
\hspace*{1mm}
\centerline{(b)}  \medskip
\end{minipage}	
\vspace*{-7mm}  
\caption{
\fontsize{9.2}{11.1}\selectfont
Block diagram of an NNSVS SVS system: (a) phonetic timing prediction and (b) waveform synthesis.
}
\vspace*{-4mm}
\label{fig:overview}
\end{figure}

Figure~\ref{fig:overview} presents an overview of an SVS system built using NNSVS. 
The phonetic timing prediction consists of time-lag and duration models, whereas the waveform synthesis consists of an acoustic model and a vocoder.

\subsection{Time-lag model}
\label{subsec:timelag}

Depending on the singing style, human vocals often deviate from the start timings of musical notes.
A time-lag model predicts those note-level timing deviations given the note-level\footnote{It is possible to use phone-level features, but we currently use note-level features for simplicity.} musical score features~\cite{saino2006hmm}.
We provide the same functionality as the latest Sinsy~\cite{hono2021sinsy}, which models the time-lags using mixture density networks (MDNs)~\cite{Bishop94mixturedensity}.

\subsection{Duration model}
\label{subsec:duration}

A duration model predicts phone durations given the phone-level musical score features.
As in the time-lag model, we employ an MDN-based duration model~\cite{hono2021sinsy}.
By combining the predicted time-lags and phone durations, the post-processing module in \figref{fig:overview} (a) computes normalized phone durations so that the sum of the predicted durations equals to the note durations~\cite{hono2021sinsy}.
After the post-processing, phone-level score features are converted to frame-level ones and used as the input of the acoustic models.

\subsection{Acoustic model}
\label{subsec:acoustic}

An acoustic model predicts frame-level acoustic features from frame-level score features\footnote{
We could support phone-level features as input for joint optimization of the duration and acoustic models, but we have not implemented it yet.
}.
We describe the details of several important implementations below.

\vspace{-1mm}
\subsubsection{Sinsy-based model}
\vspace{-1mm}
\label{sssec:sinsy}

We provide an implementation that resembles Sinsy's acoustic model~\cite{hono2021sinsy}.
The architecture of Sinsy's acoustic model consists of three fully connected layers, three one-dimensional convolution layers with batch normalization~\cite{ioffe2015batch} and ReLU activations~\cite{nair2010rectified}, followed by two bi-directional long short-term memory networks~\cite{hochreiter1997long} and a projection layer.
As the target acoustic features, vocoder parameters (e.g., MGCs and $F_0$) and additional vibrato parameters (i.e., binary vibrato flags, amplitude, and speed~\cite{nakano2006automatic}) are used.
For robust $F_0$ prediction, Sinsy adopts a pitch normalization technique: predicting the residual log-$F_0$ based on the input note pitch.

To avoid out-of-tune pitch issues, several pitch correction algorithms have been proposed~\cite{hono2021sinsy}.
Our toolkit provides one of Sinsy's pitch correction algorithms: a pitch correction method that imposes a prior distribution of pitch based on the note pitch information.

\vspace{-1mm}
\subsubsection{Multi-stream models}
\vspace{-1mm}
\label{sssec:multistream}

DNN-based acoustic models such as the one in Sinsy jointly predict $F_0$ and other spectral features.
However, it has been found that a DNN tends to prioritize higher dimensional spectral features over $F_0$~\cite{wang2018investigating}.
\rredit{To address this problem}, we provide multi-stream architectures that model each feature stream separately.
This multi-stream design enables flexible and fine-grained control for modeling different features. 
\rredit{Furthermore, to encourage predictions of multiple networks to be coherent, we implemented functionality that conditions the predictions of one network to the input of another, similar to the neural parametric singing synthesizer~\cite{blaauw2017neural}. 
In particular, we found that conditioning log-$F_0$ to the spectral feature prediction models was beneficial.
}

Our toolkit provides several generic implementations that can be used with the multi-stream architecture: convolutional neural networks (CNNs), recurrent neural networks (RNNs), and advanced architectures such as that of Sinsy and \rredit{a} duration-informed autoregressive model based on the Tacotron~\cite{okamoto2019tacotron}.
Furthermore, we provide implementations specifically designed for $F_0$ prediction.

\vspace{-1mm}
\subsubsection{Autoregresssive $F_0$ models}
\vspace{-1mm}
\label{sssec:arf0}

Modeling $F_0$ is the key to achieving expressive and natural SVS. 
Although Sinsy models the dynamic characteristics of the $F_0$ contour by an explicit vibrato modeling, 
we provide alternative implementations based on autoregressive models, of which the effectiveness has been confirmed in \rrredit{text-to-speech}~\cite{wang2018autoregressive} and SVS~\cite{blaauw2017neural,yi2019singing}.
As demonstrated in \secref{ssec:subeval}, autoregressive $F_0$ models can \rredit{generate a more natural voice without explicitly modeling vibrato}.

Inspired by Sinsy~\cite{hono2021sinsy} and XiaoiceSing~\cite{lu2020xiaoicesing}, we incorporate residual log-$F_0$ modeling within the autoregressive models.
The choice of detailed architecture may be arbitrary, but we found that a Tacotron-based RNN works well~\cite{okamoto2019tacotron}.

\subsection{Vocoder}

NNSVS supports WORLD~\cite{Morise2016WORLDAV} as a signal processing-based vocoder and uSFGAN~\cite{yoneyama22_interspeech} as a neural vocoder.
WORLD can be used to achieve reasonably good-quality SVS,
whereas neural vocoders are generally preferred to achieve better sound quality.

Note that we also support various neural vocoders based on generative adversarial networks (GANs)~\cite{goodfellow2014generative} such as Parallel WaveGAN~\cite{yamamoto2020parallel} and HiFi-GAN~\cite{kong2020hifi}. 
However, we found that uSFGAN archived a better tradeoff between quality and pitch robustness.

\section{Experimental evaluations}
\label{sec:exp}

\subsection{Database}
    
To evaluate the performance of NNSVS, we used Namine Ritsu's publicly available database~\cite{2020ritsu_enunuv2}.
The database contains 110 songs recorded by a single Japanese singer. 
Specifically, it includes 4.35 hours of singing data with timings, phonetic, and musical context annotations.
We split the 110 songs using a ratio of 100/5/5 for the training, validation, and test sets. respectively. 
We also split each song into small segments based on the rest notes in the musical scores.
Note that we selected test songs to cover a wide \rrredit{range} of \rredit{note} pitches: the lowerest and highest notes of the test songs were D\#4 (155.6~Hz) and A5 (880~Hz), \rredit{whereas those of the training data were D\#3 (146.8~Hz) and B5 (987.8~Hz).}
The audio signals were sampled at 44.1 kHz and downsampled to 24 kHz. Each audio was normalized to -26 dB.

At the pre-processing stage, we extracted 82-dimensional score features (e.g., phone identity, note pitch, and note duration) for the time-lag and duration models. Additional four-dimensional coarsely coded positional features~\cite{zen2015unidirectional} were used for the acoustic models.

\subsection{Model details}


\begin{table*}[!t]   
\vspace{-2mm}
\begin{center}         
\caption{
\fontsize{9.2}{11.1}\selectfont
Naturalness MOS test results with 95\% confidence intervals. 
MEL denotes mel-spectrogram.
A/S in the system column represents that the speech samples were generated by the extracted acoustic features. Otherwise, samples were generated by the input musical score. 
Bold font represents the best score in all SVS systems.
}
\vspace{1mm}
\label{tab:mos}
\scalebox{0.96}{
{\small        
\begin{tabular}{lllllc}
\Xhline{2\arrayrulewidth}
\multirow{2}*{System} & Acoustic & Mult-stream & Autoregressive & \multirow{2}*{Vocoder} &  \multirow{2}*{MOS$\uparrow$} \\
 & Features & Architecture & Streams & & \\
\hline
Sinsy ~\cite{hono2021sinsy} & MGC, LF0, VUV, BAP & No & - & hn-uSFGAN & $2.64 \pm 0.12$ \\
Sinsy (w/ pitch correction)~\cite{hono2021sinsy} & MGC, LF0, VUV, BAP & No & - & hn-uSFGAN & $2.84 \pm 0.11$ \\
Sinsy (w/ vibrato modeling)~\cite{hono2021sinsy} & MGC, LF0, VUV, BAP, VIB & No & - & hn-uSFGAN & $2.99 \pm 0.11$ \\
\hline
Muskits RNN~\cite{shi22d_interspeech} & MEL & No & - & HiFi-GAN & $2.22 \pm 0.11$ \\
DiffSinger~\cite{liu2021diffsinger} & MEL, LF0, VUV & Yes & -  & hn-HiFi-GAN & $2.90 \pm 0.11$ \\
\hline
NNSVS-Mel~v1 & MEL, LF0, VUV & Yes & -  & hn-uSFGAN & $3.51 \pm 0.11$ \\
NNSVS-Mel~v2 & MEL, LF0, VUV & Yes & LF0  & hn-uSFGAN & $3.58 \pm 0.11$ \\
NNSVS-Mel~v3 & MEL, LF0, VUV & Yes & MEL, LF0  & hn-uSFGAN & $2.58 \pm 0.11$ \\
\hline
NNSVS-WORLD~v0~\cite{2020ritsu_enunuv2} & MGC, LF0, VUV, BAP & No & -  & WORLD & $3.28 \pm 0.10$ \\
NNSVS-WORLD~v1 & MGC, LF0, VUV, BAP & Yes & -  & hn-uSFGAN & $3.21 \pm 0.12$ \\
NNSVS-WORLD~v2 & MGC, LF0, VUV, BAP & Yes & LF0 & hn-uSFGAN & $3.35 \pm 0.11$ \\
NNSVS-WORLD~v3 & MGC, LF0, VUV, BAP & Yes & MGC, LF0  & hn-uSFGAN & $3.60 \pm 0.11$ \\
NNSVS-WORLD~v4 & MGC, LF0, VUV, BAP & Yes & MGC, LF0, BAP  & hn-uSFGAN & \textbf{3.86 $\pm$ 0.10} \\
\hline
hn-HiFi-GAN (A/S) & MEL, LF0, VUV & - & - & hn-HiFi-GAN & $3.72 \pm 0.11$ \\
hn-uSFGAN-Mel (A/S) & MEL, LF0, VUV & - & - & hn-uSFGAN & $4.19 \pm 0.09$ \\
hn-uSFGAN-WORLD (A/S) & MGC, LF0, VUV, BAP & - & - & hn-uSFGAN & $4.19 \pm 0.09$ \\
\bottomrule
Recordings & - & - & - & - & $4.39 \pm 0.08$ \\
\Xhline{2\arrayrulewidth}
\end{tabular}}    
}
\end{center}         
\vspace*{-7mm}
\end{table*}

\vspace{-1mm}
\subsubsection{Baselines}
\vspace{-1mm}
\label{sssec:baselines}

As baseline systems, we used our implementations of Sinsy~\cite{hono2021sinsy}, Muskits's RNN-based SVS~\cite{shi2021sequence}, and the recently proposed DiffSinger~\cite{liu2021diffsinger}. We used the same MDN-based time-lag and duration models in all systems except for Muskits and DiffSinger.
In Muskits and DiffSinger, time-lag models were not used, and the duration models were jointly trained with the acoustic models.

We implemented three variants of Sinsy's acoustic model: 1) a simplified version of Sinsy without vibrato modeling, 2) Sinsy with a pitch correction algorithm, and 3) Sinsy with pitch correction and vibrato modeling. 
The detailed model architecture follows that of Sinsy~\cite{hono2021sinsy}. 
As the acoustic features, the systems use WORLD-based 65-dimensional acoustic features containing 60-dimensional MGCs, continuous log-$F_0$ (LF0), \rredit{VUVs}, and three-dimensional BAP. 
The frame shift was set to 5 ms.
For explicit vibrato modeling, three-dimensional vibrato parameters (i.e., binary flags, amplitude, and speed; denoted as VIB) were additionally used.
We did not use dynamic features as we found them to be less useful.
During synthesis, a global variance-based post-filter \rredit{was applied to the MGCs} to alleviate over-smoothing issues~\cite{toda2007speech}.

For Muskits, we used the officially provided recipe to train the RNN-based SVS~\cite{shi2021sequence}. 
The system uses syllable-level score features and predicts durations together with an 80-dimensional mel-spectrogram.
\rredit{A HiFi-GAN vocoder was used to generate waveforms from the mel-spectrogram}~\cite{kong2020hifi}.

As for DiffSinger, we used the MIDI B-version of the official source code~\cite{diffsinger}. 
The system consists of three components: 1) a mel-spectrogram prediction \rredit{based on a denoising diffusion probabilistic model~\cite{ho2020denoising}}, 2) $F_0$/VUV prediction from the mel-spectrogram, and 3) \rredit{a HiFi-GAN vocoder with harmonic-plus-noise mechanism}~\cite{wang2019neural} (referred to as hn-HiFi-GAN).
We trained each model on \rredit{Namine Ritsu's} database for a fair comparison with other SVS systems.

\vspace{-1mm}
\subsubsection{NNSVS}
\label{sssec:nnsvs}
\vspace{-1mm}

For the NNSVS systems, we used two types of SVS systems that use mel-spectrogram and WORLD-based features, respectively.
The WORLD and mel-spectrogram features contain four (i.e., [MGC, LF0, VUV, BAP]) and three (i.e., [Mel-spectrogram, LF0, VUV]) feature streams, respectively.
Note that the details of WORLD features, time-lag model, and duration model are the same as those \rredit{described in \secref{sssec:baselines}.}
We trained several multi-stream acoustic models with non-autoregressive and autoregressive models for each feature type, as listed in \tabref{tab:mos}.
For the architecture of the multi-stream models, we adopted the Sinsy architecture for non-autoregressive modeling and the duration-informed Tacotron~\cite{okamoto2019tacotron} for autoregressive modeling.
We modeled the VUV feature stream by the Sinsy's non-autoregressive architecture \rredit{in all systems}.
\rredit{For the multi-stream models, LF0 was conditioned on the input to the spectral feature prediction models (i.e., models for predicting the MGCs, BAP, and mel-spectrogram) and the VUV prediction model.}
\rpostedit{In addition, NNSVS-WORLD~v4 used MGCs as conditioning for predicting VUV.}
\rredit{Note that we did not use pitch correction methods for NNSVS.}
During synthesis, \rredit{global variance post-filters were used for the MGCs and mel-spectrogram.}
For the neural vocoder architecture, we used the harmonic-plus-noise uSFGAN (hn-uSFGAN)~\cite{yoneyama22_interspeech}.
We trained two hn-uSFGAN vocoders for the mel-spectrogram and WORLD features. Then, the vocoders were used for NNSVS and Sinsy.
All the acoustic models and vocoders of NNSVS were trained for 100 epochs and 600 K steps, respectively.
More details of the model architecture, training setups, and hyperparameters can be found in our GitHub repository\footnote{
\url{https://github.com/nnsvs/nnsvs}
}.

To evaluate the recent improvements of NNSVS, we included a publicly available pre-trained SVS model (\redit{NNSVS-WORLD v0}) that was created with an earlier version of NNSVS (November 2021)~\cite{2020ritsu_enunuv2}.
The acoustic model was based on the single-stream architecture, and it consisted of stacks of six one-dimensional CNNs with residual connections, followed by an MDN layer.
The time-lag and duration models were also based on MDNs.

\subsection{Subjective evaluation}
\label{ssec:subeval}

We performed MOS tests using a five-point scale for evaluation.
Eighteen native Japanese speakers were asked to judge the quality of the singing \rrredit{voice} samples.
For the tests, five short segments of 3 to 20 seconds were randomly selected for each of the five test songs. 
In total, 25 samples for each SVS system were evaluated.
We also evaluated the synthesized samples generated by the extracted acoustic features using the hn-HiFi-GAN and hn-uSFGAN vocoders.

\tabref{tab:mos} shows the MOS test results, of which the trends can be summarized as follows: 
\begin{inparaenum}[(1)]
    \item Sinsy with explicit vibrato modeling achieved the best score among the three Sinsy variants, which confirms the importance of modeling the dynamic characteristics of $F_0$.
    \item All NNSVS systems except for the NNSVS-Mel~v3 outperformed the Sinsy, Muskits, and DiffSinger baseline systems.
    \item The source-filter-based hn-uSFGAN performed significantlly better than hn-HiFi-GAN.
    \item The autoregressive architecture for the WORLD-based multi-stream models improved the naturalness of the SVS. 
    In particular, NNSVS-WORLD~v4 achieved the best score of all the SVS systems (3.86), demonstrating the effectiveness of multi-stream and autoregressive architectures for \rredit{modeling $F_0$ and spectral features}.
    \item Comparing NNSVS-WORLD~v4 and NNSVS-WORLD~v0, we confirmed that we have obtained substantial performance improvements over the earlier version of NNSVS.
\end{inparaenum}
 
We observed that NNSVS-Mel~v3 tended to generate unstable output for low- and high-pitched voices.
We hypothesize that NNSVS-Mel~v3 obtained a lower score than \rpostedit{the} other NNSVS systems\rpostedit{, such as NNSVS-WORLD v3, because of the exposure bias issues}. 
\rpostedit{In particular, the mel-spectrogram was more challenging to model than the MGC and BAP since the mel-spectrogram is more complex and entangled representation (i.e., it contains F0 and harmonic/aperiodicity envelopes), while MGC and BAP are disentangled~\cite{blaauw2017neural,kim2018korean}}.
We also note that even though DiffSinger generated a singing voice with higher fidelity than other systems, it often generated \rpostedit{unnatural pitches such as discontinuous F0 and unstable vibrato}, especially for dynamic voices.
\rredit{We encourage readers to listen to the singing voice samples provided on our demo page\footnote{\url{https://r9y9.github.io/projects/nnsvs/}}.}

\section{Conclusion}
\label{sec:conclusions}

This paper described the design of NNSVS, an open-source toolkit for SVS research. 
Our toolkit provides implementations of Sinsy and many new features such as multi-stream models, autoregressive $F_0$ models, and neural vocoders \rredit{based on uSFGAN}.
Experimental results demonstrated that our best system \rredit{significantlly} outperformed the Sinsy, Muskits, and DiffSinger baseline systems. 
Future work includes adding more advanced architectures based on variational auto-encoders, diffusion models, and GANs.
We also plan to work on end-to-end models.

\hfill \break
\noindent\textbf{Acknowledgments}:
This work was partly supported by JST CREST Grant Number JPMJCR19A3.

\vfill\pagebreak


\section{References}
{
\setstretch{0.895}
\printbibliography
}

\end{document}